\documentclass[useAMS,usenatbib]{mn2e}

\usepackage{graphicx}

\title[Local effects in astrometric binary
orbits]{Local effects in astrometric binary orbits: Perspective transformation and light--travel time}
\author[J.-L. Halbwachs]{J.-L. Halbwachs$^{1}$\thanks{E-mail:
halbwachs@astro.u-strasbg.fr}\\
$^{1}$Observatoire Astronomique de Strasbourg (UMR 7550),
   11 rue de l'Universit\'{e}, F--67\,000 Strasbourg, France}
\begin{document}

\date{Accepted . Received 2008 ; in original form }

\pagerange{\pageref{firstpage}--\pageref{lastpage}} \pubyear{2009}

\maketitle

\label{firstpage}

\begin{abstract}
The next generation of astrometric instruments will reach accuracies 
deserving new treatments. In order to get astrometric parameters achieving the
precision permitted by the measurements, it will be necessary to take into account 
effects that were neglected until the present time. 
Two effects concerning the orbital elements of binary stars are considered hereafter:
the former is the local perspective (LP) effect, which is due to the variation of the distance and of the orientation of the orbital plane during the observation time span; the equations describing this effect are derived for the first time. The latter effect is
the light--travel time (LTT), which is also related to the orientation of the orbital
plane, and which is as efficient as the preceding one. 
Taking these effects into account would allow to find the ascending nodes of the orbits,
and lead to orbital elements more accurate than when they are ignored. 

It is derived from simulations that, at a distance of 5~pc, and assuming velocities
typical of Pop.I stars, the position of the right ascending node could be derived for a few simulated unresolved binaries when the astrometric measurements have errors around 1~$\umu$as.
For the resolved brown dwarf binary 2MASS J07464256 +2000321, it appears that ignoring the 
LP effect would result in underestimating the masses
of the components by 14 per cent of the errors as soon as the astrometric errors are around 20~$\umu$as for each measurement.
However, a `degenerate LP solution', taking into
account the variation of the semi-major axis when the distance is varying, should provide
reliable masses when the measurement errors are larger than 1 or 2~$\umu$as. A few binaries
in the {\it Gaia} program could deserve a degenerate LP solution, whereas a the complete LP+LTT 
solution could be justified for resolved binaries observed with {\it SIM}.

\end{abstract}


\section{Introduction}

Thanks to the progresses in spatial astrometry, the errors of measurements of the
positions of stars are gradually decreasing: the errors of the along--scan
measurements were a few milli--arcseconds in the {\it Hipparcos} project \citep{hip}, but the {\it Gaia}
satellite \citep{perryman} is expected to achieve 100 times better \citep{Mignard}. This limit will still be pushed beyond 1~$\umu$as when interferometry with large baseline will be feasible in space. It is expected that,
with a 9-m baseline, {\it SIM} could already get an accuracy of only 0.6~$\umu$as for a single measurement \citep{SIM}.

These important improvements will not only lead to more accurate results. They make also necessary to take into account effects that were never considered in the past. A first example was the correction for perspective acceleration, that was introduced for a few high-velocity single stars in the {\it Hipparcos} catalogue. The radial velocities (RV) of these stars were used for that purpose, but \citet{Dravins}
have shown that the RV may also be derived from this effect.
\citet{Klioner03} developed a relativistic model taking into account the
light deflection by solar system objects.
\citet{Klioner00} and \citet{Mignard} outlined that some basic parameters as the
trigonometric parallax, the proper motion and the RV deserve 
to be defined more precisely than they are at present; the definition of the RV was discussed by \citet{linde03}. Concerning binary stars,
\citet{Angla06} have shown that the variations of the
light--travel time (LTT) coming from the orbital motion may change the apparent position
of the component. Since this effect is related to the radial displacement of the component, they inferred that it could be used to dissipate the ambiguity of the position of the ascending node of the orbit. Another effect which could be used for node identification is
the modification of the orientation of the orbit during the time span of the astrometric program. For sake of brevity, it is called `local perspective' (LP) hereafter. In order to avoid confusion, the
perspective effects concerning the barycentre of binary systems, but also single stars (i.e. perspective acceleration and
change in the parallax), are called `barycentric perspective' hereafter.

In the course of a high-precision astrometric project, taking perspective and LTT effects into account could be
necessary for two reasons. First of all, when the effects are relevant, the quality of the solution
is improved. The astrometric and orbital parameters are then more accurate, and,
even more important, the solution itself will look more `acceptable'. The decision to consider
a model as adequate is usually based on the goodness-of-fit (GOF) of the solution, which may, for instance, follow the normal distribution ${\cal N}(0,1)$ when the model is the right one. Therefore,
a large GOF leads to the rejection of the model, and to the search of another one. In the course of a large astrometric survey, like {\it Hipparcos}, objects which are not satisfactorily fitted
eventually got a `stochastic' solution, which is of rather poor interest. Therefore,
a binary model neglecting perspective or LTT effects could change actual binary stars in stochastic objects, losing the orbital elements. Aside from giving the right classification with accurate parameters, the second advantage of giving a precise solution is the possibility to get more informations about the
object. For single stars, as mentioned above, taking into account perspective acceleration can lead to an estimation of the RV. For binaries, a consequence of the additional effects could be the complete derivation of the orientation of the orbit in space. This is investigated hereafter.

Calculations were performed in order to answer three questions: 
What is the smallest astrometric error still permitting to neglect LP and LTT without altering dramatically the GOF for unresolved binaries? What is the largest astrometric error permitting to derive the precise ascending node thanks to LP and LTT? What is the smallest astrometric error still permitting to neglect LP and LTT without altering the masses of
resolved binary components?
The basic equations of LP are derived in Section~\ref{sec:LP}, and it is shown how to select the right
ascending node thanks to this effect. A method for selecting the ascending node from LTT is presented in Section~\ref{sec:LTT}. In Section~\ref{sec:relat}, it is verified that the relativistic effects, which were
neglected in the calculations, are much smaller than LP and LTT effects for nearby binaries with large
semi--major axes. The parameters of simulated unresolved astrometric binaries were derived in Section~\ref{sec:simulation}, and the ascending nodes were searched. Resolved binaries are treated
in Section~\ref{sec:resolved}. Section~\ref{sec:conclusion} is the conclusion.


\section{The local perspective effect}

\label{sec:LP}

\subsection{General transformations}
\label{sec:local:trans}

\begin{figure}
\begin{center}
\includegraphics[clip=,height=3.1 in]{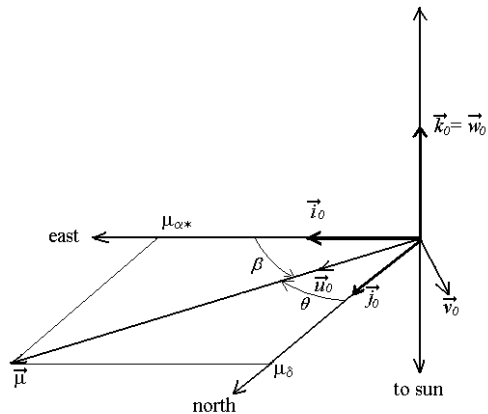}
\end{center}
 \caption{
The first coordinate transformation in computing the local perspective effect: 
Rotation of angle $\beta$ of the local equatorial reference frame around the radial vector for aligning the rectangular 
coordinates with respect to the proper motion. The angle $\theta$ is the position angle of the proper
motion; it is measured in the anticlockwise direction, but inside the celestial sphere.
}
\label{fig:LPtrans1}
\end{figure} 

\begin{figure}
\includegraphics[clip=,height=4.1 in]{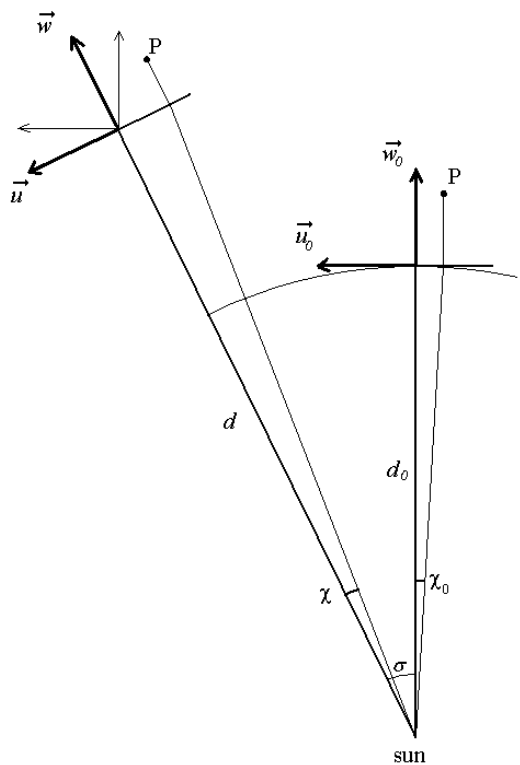}
 \caption{
Perspective transformation when the
barycentre of the object is moving in the 3-D proper motion direction. The reference system is
rotating with the $\sigma$ angle ($\sigma = \mu \; t$, or $\sigma = \mu \; t \times (1-\mu_r t)$
when perspective acceleration is taken into account), and the angular dimensions are
transformed by a factor $d/d' = 1-\mu_r t$.
}
\label{fig:LPtrans2}
\end{figure}

We want to derive the angular coordinates of binary components in a tangential reference frame
moving with the barycentre. At the beginning, the components are located in the `local equatorial reference system'
of the object, which is defined as a trihedron having the origin on the barycentre of the binary, and with the vectors $(\vec{i}, \vec{j}, \vec{k})$ oriented as follows:
$\vec{k}$ is prolongating the line of sight, $\vec{i}$ is oriented toward east,
and $\vec{j}$ is oriented to north, in order to complete the direct trihedron 
(figure~\ref{fig:LPtrans1}). 

The angular coordinates are lengths divided by the distance
from the Sun.
At the epoch $t=0$, the object is moving in space with the angular velocity ($\mu_{\alpha *}, \mu_\delta, \mu_r$). The third coordinate, 
$\mu_r$ is the `radial proper motion', which is always expressed in radians per time unit. It may be derived from
the RV, $v_r$, and from the parallax, $\varpi$, using the equation $\mu_r = v_r \varpi \times (1 - v_r/c)$, where $v_r$ is in AU per time unit and $\varpi$ is in radians; $c$ is the speed of light, in the same unit as
$v_r$.

We consider a point with a fixed position in an inertial reference
frame attached to the barycentre (for instance: the periastron of an astrometric orbit).
When the binary is moving, the angular coordinates of this point in a local reference system
attached to the barycentre are resulting from a set of consecutive transformations:

\begin{enumerate}
\item
A rotation about the radial axis, $\vec{k_0}$, for aligning the first coordinate axis along the proper motion in the ($\vec{i_0}$, $\vec{j_0}$) tangential
plane (figure~\ref{fig:LPtrans1}). The rotation angle $\beta$ is derived from 

\[
\cos \beta = \mu_{\alpha *}/\mu,
\]
\[
\sin \beta =\mu_\delta/\mu.
\]

Let ($\vec{u_0}, \vec{v_0}, \vec{w_0}$) be the new reference system, and let $\chi_0$, $\psi_0$ and $\zeta_0$ be the coordinates of the point in this system.

\item
The transformation of the ($\vec{u_0}, \vec{v_0}, \vec{w_0}$) reference system in 
($\vec{u}, \vec{v}, \vec{w}$), due to the displacement of the star. This transformation consists in
a rotation about $\vec{v_0}$ of an angle $\sigma$, and in a correction of the angular coordinates when
the distance of the barycentre, $d_0$, is changed in $d$ (figure~\ref{fig:LPtrans2}).
$\sigma$ is the motion of the barycentre due to the proper motion, and, in a first order approximation, $d_0/d = 1 - \mu_r t$.
Therefore, the coordinates in the ($\vec{u},  \vec{v}, \vec{w}$) system are:

\[
\left(
\begin{array}{c}
\chi \\
\psi \\
\zeta
\end{array}
\right) \; = \;
(1 - \mu_r t) \; \left(
\begin{array}{ccc}
\cos \sigma & 0 & -\sin \sigma \\
0 & 1 & 0 \\
\sin \sigma & 0 & \cos \sigma
\end{array}
\right)
\; 
\left(
\begin{array}{c}
\chi_0 \\
\psi_0 \\
\zeta_0
\end{array}
\right) 
\]

where $(\chi_0, \psi_0, \zeta_0)$ refers to the ($\vec{u_0}, \vec{v_0}, \vec{w_0}$) frame. 

\item

A rotation about the radial axis $\vec{w}$ must still be applied in order to get the coordinates of the point in the most suitable
reference frame for the forthcoming calculations. 
In Section~\ref{sec:abscissa}, the motion of the binary is projected on a unique equatorial tangential plane, corresponding to a point arbitrarily chosen. This plane is called the `average' equatorial tangential plane hereafter; for simplicity, we chose, for reference frame, the local equatorial tangential frame
of the beginning, ($\vec{i_0}, \vec{j_0}$). Therefore, we need, for each barycentric position, an orientation of the ($\vec{i}, \vec{j}$) axes such that the projections of these axes on the average equatorial tangential plane are parallel to ($\vec{i_0}, \vec{j_0}$), respectively. Since the projection of the motion of the barycentre is a straight line on the average plane,
this condition is achieved by applying to the ($\vec{u}, \vec{v}, \vec{w}$) frame a rotation with the angle $-\beta$. The ($\vec{i}, \vec{j}, \vec{k}$) frame thus obtained is called the `local standard reference system' hereafter.

In a first order calculation, $\sin \sigma = \mu \; t$, $\cos \sigma =1$, and the products between
the angular motions, such as $\mu_r t \times \mu_{\alpha *} t$, are negligible. Therefore,
the ($x, y, z$) coordinates
in the  ($\vec{i}, \vec{j}, \vec{k}$) reference frame are given by the equations:
 
\begin{equation}
x = (1 - \mu_r t) \; x_0 - [\mu_{\alpha *} t ]_{rad} \; z_0 
\label{eq:cooper1}
\end{equation}
\begin{equation}
y = (1 - \mu_r t) \; y_0 - [\mu_{\delta} t ]_{rad} \; z_0
\label{eq:cooper2}
\end{equation}
\begin{equation}
z = (1 - \mu_r t) \; z_0 + [\mu_{\alpha *}t]_{rad} \; x_0 + [\mu_{\delta} t]_{rad} \; y_0
\label{eq:cooper3}
\end{equation}

\noindent
where `$[ \; ]_{rad}$' means that the angle between the brackets is converted in radians 
($\mu_r t$ is also in radians, according to the definition of $\mu_r$ above).

An anonymous referee pointed out that the local standard reference system ($\vec{i}, \vec{j}, \vec{k}$) is not
a local
equatorial reference system, contrarily to the ($\vec{i_0}, \vec{j_0}, \vec{k_0}$) frame of the beginning of the calculation. 
The coordinates in the local equatorial reference system are not used in the derivation of the
orbital elements with the method of the average equatorial tangential plane, in Section~\ref{sec:simulation}, but they may be relevant for high-velocity binary stars, when this
method cannot be applied. For that reason, they are derived hereafter.

For getting
the coordinates in the local equatorial reference system
of the barycentre, it would still be necessary to add a rotation about $\vec{k}$ with an angle equal to the rotation of
the north, but in the opposite direction. If $\Delta \theta$ is the correction to be added to the
position angles measured inside the celestial sphere, as in the usual definition (figure~\ref{fig:LPtrans1}), then, the angle of the rotation to apply to
the ($\vec{i}, \vec{j}, \vec{k}$) reference frame for getting an equatorial frame is also $\Delta \theta$.
It comes from the basic formulae of spherical trigonometry
\citep{vandeKamp} that $\Delta \theta$ is given by the exact equation:

\[
\tan \Delta \theta = \frac{\sin \sigma \tan \delta \tan \theta + (1 - \cos \sigma) \sin \theta } {
1 - \sin \sigma \tan \delta  \cos \theta - (1 - \cos \sigma) \cos^2 \theta}
\]

When $\sigma$ is small, the equation above may be simplified assuming $(1 - \cos \sigma)=0$. If the
binary is not close to a pole, it may also be assumed that $\sigma \tan \delta  \cos \theta \ll 1$. Using the same approximations as above, the referee obtained:

\[
\Delta \theta = [\mu_{\alpha *}t]_{rad} \tan \delta
\]

Therefore, applying a rotation with a small $\Delta \theta$ angle leads to the coordinates

\[
x_1 = x + [\mu_{\alpha *}t]_{rad} \tan \delta \; y_0
\]
\[
y_1 = y - [\mu_{\alpha *}t]_{rad} \tan \delta \; x_0
\]

\noindent
in the local equatorial tangential plane of the barycentre.

\end{enumerate}


\subsection{Application to an orbit}

\subsubsection{The ascending node problem}
\label{sec:ascendingNode}

Two different sets of parameters are commonly used to entirely describe an orbit in space. Both are including
the period, $P$, the eccentricity, $e$, and the periastron epoch, $T_0$. These parameters may be completed with the angular semi-major axis, $a$, and the Campbell's angles, which are:
the inclination, $i$, the position angle of the ascending node, $\Omega$, and the so-called longitude of periastron, $\omega$, which is measured from the ascending node; the definitions of these angles may be found in any double star handbook \citep{Binnen,Heintz}. Hereafter, the semi-major axis refers to the astrometric orbit of the primary component, which should not be confused with the relative `visual' orbit of a resolved binary system.
The nodes are the intersections between the orbit and the ($\vec{i}, \vec{j}$) plane; the
ascending node is the one where the star is moving away from the Sun. It is worth noticing that the positions of the nodes are derived from astrometric data, but that the ascending node is usually not known when no RV measurement is available. For an astrometric orbit, $\Omega$ and $\omega$ refer then to the node with $\Omega$ between 0 and $\pi$. However, when the right ascending node is eventually found, it may be the other one, and $\Omega$ and $\omega$ are then changed in $\Omega + \pi$ and $\omega+\pi$, respectively.

The second set of parameters which are frequently used are the Thiele-Innes elements ($A,B,F,G$). These elements are functions of $a$ and of the Campbell's angles, so it is
possible to go from one system to the other and vice-versa \citep{Binnen}.
However, changing $\Omega$ and $\omega$ in $\Omega + \pi$ and $\omega+\pi$ doesn't change
($A,B,F,G$). Reciprocally, changing ($A,B,F,G$) in ($a, i, \Omega, \omega$) leads again to the ambiguity in the choice of the node.

The Thiele-Innes elements are completed with $C$ and $H$, which are defined by the following
equations:

\begin{equation}
C= a \sin i \sin \omega
\label{eq:C}
\end{equation}
\begin{equation}
H= a \sin i \cos \omega
\label{eq:H}
\end{equation}

These two additional elements are not entirely redundant with the four former, since they indicate the true ascending node, when it may be fixed.  Otherwise, it is impossible to
discriminate between ($C,H$) and ($-C,-H$).


\subsubsection{Correcting the Thiele-Innes elements for LP}
\label{sec:AtBtFtGt}

The six Thiele-Innes elements,
($A, B, C, F, G, H$), are completely fixing the orbit of a binary component in space. 
In the local reference system defined in Section~\ref{sec:local:trans} above, the
direction cosines of the semi-major axis and of the semi-minor axis of the orbit are
($B/a, A/a, C/a$) and ($G/a,F/a,H/a$) respectively. The transformation of ($A,B,F,G$)
by LP is obtained by applying equations (\ref{eq:cooper1}) and
(\ref{eq:cooper2}) to these coordinates; one obtains:

\begin{equation} 
\left\{ \begin{array}{l}
A = (1 - \mu_r t) A_0 - [\mu_{\delta} t]_{rad} \; C_0  \\ 
B = (1 - \mu_r t) B_0 - [\mu_{\alpha *} t]_{rad} \; C_0 \\
F = (1 - \mu_r t) F_0 - [\mu_{\delta} t]_{rad} \;  H_0 \\
G = (1 - \mu_r t) G_0 - [\mu_{\alpha *} t]_{rad} \; H_0 
\end{array}
\right.
\label{eq:ABFGt}
\end{equation}

It is worth noticing that the elements thus obtained are not the {\it true} Thiele-Innes elements,
since the position angle of the line of nodes is not measured exactly from the local north, but from
the $\vec{j}$ vector, ie from the north at $t=0$.

The position of a binary component is derived from the ($A,B,F,G$) elements with the
classical formulae; in the ($\vec{i},\vec{j} ,\vec{k} $) coordinate system defined in 
Section~\ref{sec:local:trans}, they are:

\begin{equation}
\left\{ \begin{array}{l}
x =  B \; X \; + \; G Y \\
y = A \; X \; + \; F Y 
\end{array}
\right.
\label{eq:xy}
\end{equation}

with

\begin{equation}
\left\{ \begin{array}{l}
X(t) = \cos E(t) -e \\
Y(t) = \sqrt{1-e^2} \;  \sin E(t)
\end{array}
\right.
\label{eq:XY}
\end{equation}

\noindent
where $E(t)$ is the eccentric anomaly at the epoch $t$.


\subsubsection{Fixing the ascending node from local perspective}
\label{sec:ascNodeLP}

As explained in Section~\ref{sec:ascendingNode}, when the actual ascending node is not known, the
($A_0, B_0, F_0, G_0$) elements correspond to two sets ($C_0,H_0$): the first one
with $\omega \in [0,\pi[$ and the second one with $\omega \in [\pi, 2\pi[$. However, thanks to the perspective
effect, it is possible to discriminate between the two possibilities, using the following method: The 
($A,B,F,G$) elements at each observation epoch are derived from equation (\ref{eq:ABFGt}), and the model positions of the binary component are derived twice, assuming 
the two $\omega$. The right ascending node
will provide the solution with the smallest $\chi^2$.


\subsubsection{Size of the LP effect}
\label{sec:sizeLP}

Our aim is to estimate of the size of the LP effect under
the best conditions. For that purpose, we consider a binary at a distance of 5~pc.
In order to get estimations applying for resolved binaries as well as for unresolved binaries,
the mass ratio (mass of the secondary component divided by that of the primary one) $q=0.5$ is assumed. For resolved binaries, the semi-major axis
of the secondary orbit is then exactly twice that of the primary, and for unresolved binaries,
the magnitude difference is large enough for neglecting the difference between the position of the primary and that of the photocentre.
The semi-major axis of the primary orbit is set to 1~AU; for a
1-solar mass primary and $q=0.5$, this corresponds to a period of $\sqrt{18}=4.24$~years.
The assumed time span of the observations is 5 years, like the expected duration of 
the {\it Gaia} mission. The orbit is then entirely covered by the observations. In order to choose a reasonable proper motion, a velocity
of 25~km/s is assumed for each axis: this value is close to the standard deviation of the velocity of late-type dwarfs, in any direction \citep{LB}.

We consider now an edge--on orbit with the semi--major axis along the line of sight and the periastron
between the barycentre and the sun. Therefore, $A_0=B_0=0$, 
$C_0 = -a$;
Assuming that the star is exactly at the opposite of the periastron when $t=$5~years, one obtains $E=\pi$, and
equation (\ref{eq:XY}) leads to $X=-(1+e)$, $Y=0$. Applying equations (\ref{eq:ABFGt}) and (\ref{eq:xy}),
the coordinates of the component are: $x = -[\mu_{\alpha *} t]_{rad} \; a \; (1+e)$, and $y=-[\mu_{\delta} t]_{rad} \; a \; (1+e)$, instead of 0 when the perspective effect is absent. The offset in position in then
$\sqrt{x^2 + y^2}=7.2$~$\umu$as for a nearly circular orbit.

This estimation applies to a 1~AU semi-major axis at a distance $d=5$~pc, but it
may be adapted to other values. Since
the Thiele-Innes elements are varying as $a_{AU}/d$ and the $\mu$ coordinates
are varying as $v_{km/s}/d$ the LP effect is varying as $v_{km/s} \; a_{AU}/d^2$. Therefore, it is vanishing rapidly when distant stars are
considered.


\section{The LTT effect and the ascending node}
\label{sec:LTT}


\subsection{The LTT effect}
\label{sec:ascNodeLTT}

The light coming from a component orbiting around the barycentre of a binary system
is arriving with a delay, $t_B +\Delta t$, where $t_B$ is the delay of light coming from the barycentre, and where 
$\Delta t$ is coming from the position of the component on the orbit. For that reason, $\Delta t$ is called the
`orbital delay' hereafter. It is negative
when the component
is closer to the Sun than the barycentre, and negative otherwise.
The analytic expression of $\Delta t$ was derived by
\citet{Angla06}. For our purpose, a simple calculation neglecting the relativistic corrections is sufficient, as it will be shown in Section~\ref{sec:relat}.

The radial angular coordinate with respect to the barycentre
is derived from $C$ and $H$ with the following equation \citep{couteau}:

\begin{equation}
z = C \; (\cos E - e) \; + \; H \sqrt{1-e^2} \sin E
\label{eq:z}
\end{equation}

Therefore, the orbital delay is

\begin{equation}
\Delta t = \frac{z}{c \varpi}
\label{eq:Dt}
\end{equation}

\noindent
where $c$ is expressed in AU per time unit, and where
$\varpi$ is in the same unit as $z$, i.e. as $C$ and $H$. 
An astrometric measurement obtained at the epoch $t$ then provides the position of the component at the epoch $t - \Delta t$. This induces a shift of the position of the star which is related to the orbital motion, but also to the proper motion of the barycentre (see the Fig.~2 in Anglada-Escud\'e \& Torra). 

As for LP, the choice of the true ascending node is feasible trying the
two possibilities. Shifting from one node to the other consists in changing the sign of
$z$, and then of $\Delta t$. Again, the ascending node should provide the best fit when this
effect is visible.


\subsection{Size of the LTT effect}
\label{sec:sizeLTT}

For comparing the offset of position due to the orbital delay to the LP effect evaluated in Section~\ref{sec:sizeLP}, the same
hypotheses are assumed. With an edge-on orbit, the orbital delay is
then as large as 8.3 minutes at maximum. The corrections of position coming from
the orbital motion and from the proper motion are evaluated hereafter.


\subsubsection{The orbital motion}
The orbital velocity is given by the equation:

\[
v_{AU/y} = 2 \pi \sqrt{ \frac{{\cal M}}{a_{AU}} } \frac{q^{3/2}}{1+q}
\]

\noindent
where ${\cal M}$ is the mass of the primary star, in solar unit; ${\cal M}=1$ is assumed. With
$q=0.5$, and a period of 4.24 y, $v_{AU/y} = 1.5$. For 8.3~min, the correction is then 4.7~$\umu$as for the primary component at a
5~pc distance. When another distance or another semi-major axis (i.e. another period) are
considered, since $\Delta t$ is proportional to $a_{AU}$, this correction is
varying as $\sqrt{a_{AU}}/d_{pc}$. Again, it is also twice larger for the secondary component than
for the primary.


\subsubsection{The proper motion}

Since the primary component is observed with an orbital delay $\Delta t$, the
position of the barycentre which is used as origin of the orbital motion is no more corresponding to
the epoch $t$, but to $t-\Delta t$. The
apparent orbit is then distorted, since $\Delta t$ is varying in relation with the orbital motion.

For consistency with Section~\ref{sec:sizeLP}, a tangential velocity of $25 \sqrt{2}=35$~km/s is assumed. Therefore, 
the correction of the position for a 8.3~min delay is 24~$\umu$as for a distance of 5~pc.
This term is varying as $v_{km/s} \times a_{AU}/d_{pc}$.

It is worth noticing that the parallactic motion is not affected by $\Delta t$, since it is due to the
position of the Earth with respect to the Sun at the time of the observation.


\section{Local relativistic effects}
\label{sec:relat}

In Sections~\ref{sec:LP} and \ref{sec:LTT} above, the relativistic effects due to the displacement of the binary with respect to the barycentre of the solar system were ignored. In this Section, it is verified
that this approximation doesn't matter to the problem we are dealing with.

The linear motion of the barycentre of the binary has two consequences: the lengths along the direction of the displacement are shortened, and time is dilated. The coordinate of the
primary component along the velocity vector, $\vec{v}$, is then multiplied by $\sqrt{1-v^2/c^2}$. Therefore, with the same hypotheses as above (spatial velocity of $25 \sqrt{3} = 43$~km/s, orbital radius of 1~AU, distance of 5~pc) we get a change in  position that may be as large as $2 \; 10^{-3}$~$\umu$as. This is much smaller than the LP or LTT effects. 

The effect of time dilatation is as if, for the observer, the period of the binary is increased by
$1/\sqrt{1-v^2/c^2}$. This linear modification of time has no effect on our calculations. Anyway, it induces an amount of only 1.4~s for a 4.2 year period.

We may also consider the deflection of light by the mass of the secondary component. For a distant binary object, we use the approximate formula of light bending \citep{straumann}:

\[
\delta \phi = 4 G \frac {\cal M}{c^2 b}
\]

\noindent
where $G$ is the gravitational constant, ${\cal M}$ is the mass of the deflecting star, and $b$ is the `impact parameter' of the ray, measured from the centre of the deflecting star. Therefore, in the reference system defined in Section~\ref{sec:LP}, the deflection of the light emitted by
the primary component by the secondary component is, approximately:

\[
\delta \phi = 4 G \frac {{\cal M}_2}{c^2} \frac{z_{1|2}}{b}
\]

\noindent
where $z_{1|2}$ is the radial angular coordinate of the observed component (labelled `1') in a local reference system with component 2 as origin; therefore, deflection occurs only when $z_{1|2}>0$.
With the same hypotheses as before, we have ${\cal M}_2= 0.5$ solar mass, $z_{1|2}=3 \; {\rm AU} \; / \; 5 \; {\rm pc} =
0.6$~arcsec. The light deflection may then be as large as
around 1.3 $\umu$as when $b= 2$~solar radii. This is not negligible, but corresponds to a rather rare situation, i.e. an orbit almost perfectly edge-on. Since $\delta \phi$ is varying as $1/b$, it vanishes very fast when $b$ is increasing from a very small value. When the median value of $\pi/3$ radians is assumed for the orbital inclination, $\delta \phi \approx 0.007$~$\umu$as only. In conclusion, light bending
may be neglected in a first approximation, although it may be important for some systems.


\section{LP and LTT effects in simulated unresolved binaries}
\label{sec:simulation}

It comes from Section~\ref{sec:sizeLP} and \ref{sec:sizeLTT} that LP and LTT effects
could significantly alter the position of the photocentre of an unresolved binary when the accuracy is around 10 $\umu$as. 
However, simulations are necessary to evaluate when the signatures of these effects
should be visible in reality.

In this section, we attend to answer the two questions addressed in the introduction: 1) how does the distribution of the GOF look when LP and LTT are neglected?, and 2) When is it possible to select the actual ascending node in practice? 

\subsection{The simulated binaries and their observations}
\label{sec:hypsimu}

\subsubsection{Properties of the virtual systems}
Several hypotheses in 
\label{sec:binSimu}~\ref{sec:sizeLP} are assumed again for generating virtual
binaries: the mass of the primary component is 1 solar mass and that of the invisible companion is 0.5 solar mass. The period is 4.24 years, so that the
semi-major axis of the astrometric orbit is exactly 1~AU; the distance is 5~pc. In addition,
each coordinate of the barycentric velocity obeys a normal distribution with the standard deviation of 25~km/s, the positions of the stars are randomly distributed on the sky, and the orbits are not circular, but have eccentricities obeying the distribution of \citet{Halb}.
The orientations of the orbits are
purely at random: the
inclinations obey
a sinus distribution, and the $\Omega$ and $\omega$ angles obey constant distributions.


\subsubsection{Properties of the observations}
In order to reproduce the data coming from a rotating satellite or an interferometer, we consider one-dimensional astrometric coordinates (`abscissae' hereafter) measured
along an axis  with a varying orientation.
For a given binary system, the origin of the abscissa corresponds to a fixed position close
to the position of the barycentre at $t=0$ (Fig.~\ref{fig:wBin}). 
Virtual observations are produced assuming a 
rough approximation of the {\it Gaia}
scanning law, consisting in around 16 visibility periods (VP, hereafter) in 5 years, 
separated by a duration a bit longer than 100 days. A constant orientation is randomly
fixed for each VP, which contains 2 or 3 pairs of scans. The total number of measurements is 83 on average. Each pair corresponds to a complete rotation of the satellite in 6 hours. The second scan of the pair is coming 106~min after the first one. For any scan, an abscissa is derived, and an error is
generated from a normal distribution with a constant standard deviation, $\sigma_w$.


\subsubsection{Deriving the abscissae from the parameters of the stars}
\label{sec:abscissa}

\begin{figure}
\includegraphics[clip=,height=2.4 in]{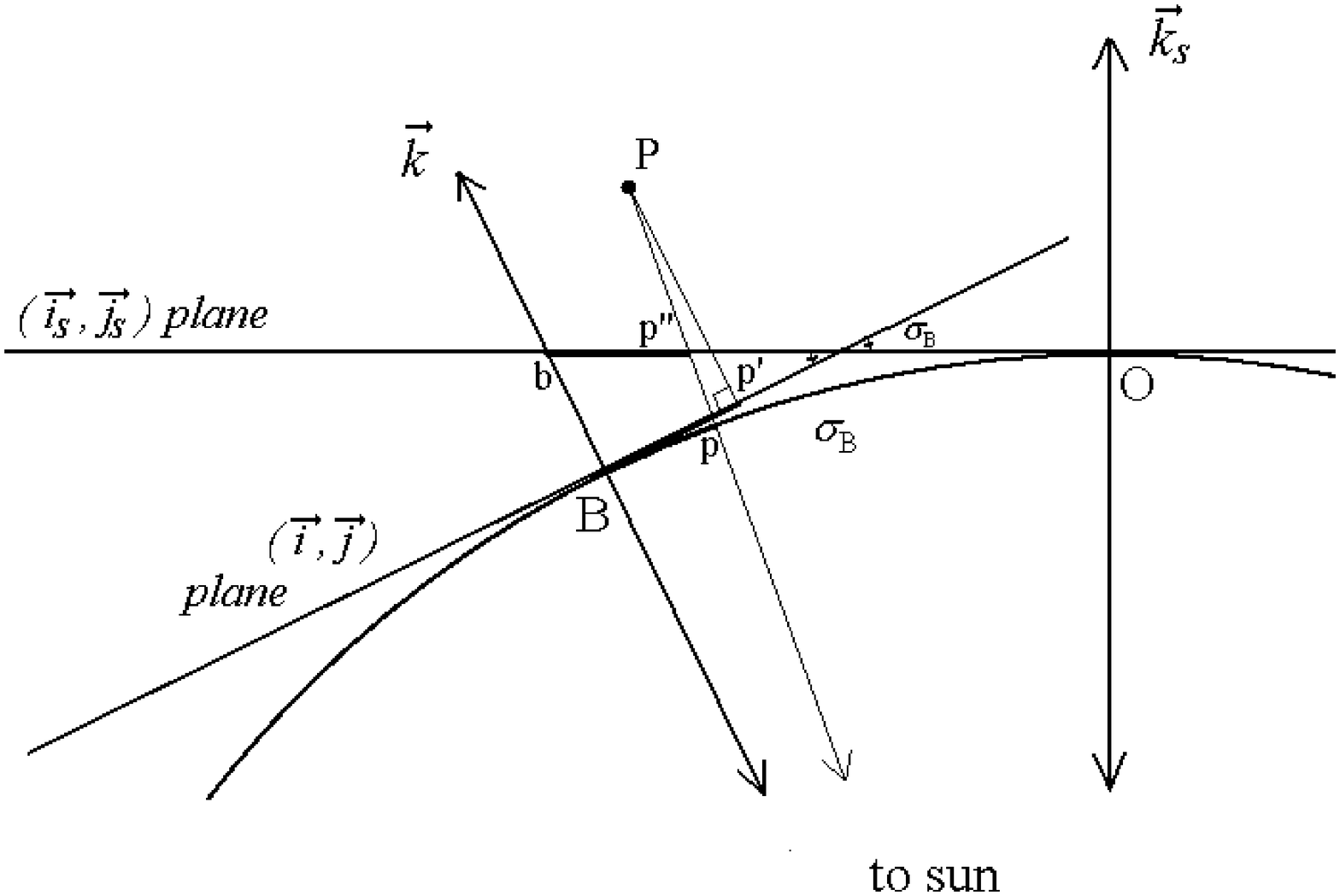}
 \caption{
Cross section of the celestial sphere, from the local standard tangential plane 
($\vec{i}, \vec{j}$) to the average equatorial tangential plane ($\vec{i_s}, \vec{j_s}$). 
{\rm B} is the barycentre of the binary as seen on the celestial sphere, and {\rm P} is a component or a
peculiar point of the orbit. The angular distance between {\rm B} and {\rm P} is the arc {\rm Bp}. It is nearly equal to  $\| {\rm B}{\rm p'} \| \approx \| {\rm b}{\rm p"} \| $, where
{\rm p'} is the projection of {\rm P} on the ($\vec{i}, \vec{j}$) plane, and $ \| {\rm b}{\rm p"} \|$
is the projection of {\rm Bp} on the average equatorial tangential plane.
}
\label{fig:tanCoo}
\end{figure} 

The easiest way to derive the parameter of a star is to consider its motion in the average equatorial tangential plane defined in Section~\ref{sec:local:trans}.
When the distance $\sigma_B$ between the origin and the barycentre of the star is small, the curvature of the
celestial sphere may be neglected, and it is assumed that $\tan \sigma_B = \sigma_B$. The coordinates of
a binary component in
the local standard tangential plane, ($\vec{i}, \vec{j}$), seen in Section~\ref{sec:local:trans} may then be directly transposed to the average equatorial tangential plane,
neglecting the tilt with the angle $\sigma_B$ between the two planes (see Figure~\ref{fig:tanCoo}); they
are just added to the coordinates of the barycentre.

However, we must verify that the curvature of the celestial sphere may really be neglected. Assuming a 5~pc distance
and a velocity as large as $2 \sigma$ on each axis
in the tangential plane, one obtain, $\mu = 3.0$~arcsec yr$^{-1}$. For half the duration of a 5~yr mission,
this implies a displacement of 7.5~arcsec. The difference between the angular distance and its projection
on the tangent plane is then only 0.003~$\mu$as. It is then quite possible to use the plane approximation, even
when astrometric errors as small as 0.1~$\mu$as are considered.

\begin{figure}
\includegraphics[clip=,height=2.4 in]{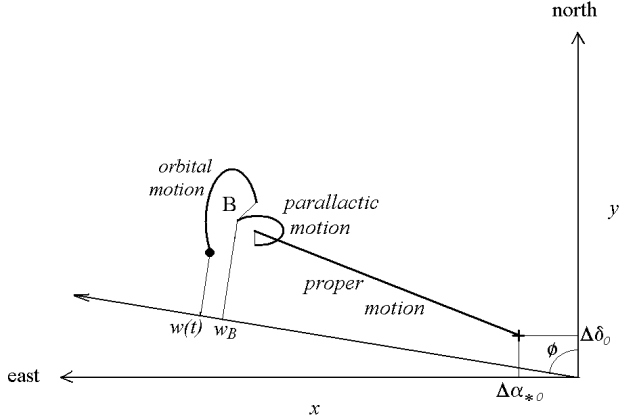}
 \caption{
The abscissa of a binary component in the average equatorial tangential reference plane. It is the projection of the heliocentric position of the barycentre at $t=0$ (indicated by the cross), {\it plus} contributions of the proper proper motion, of the parallax, and of the orbital motion of the star around B, the position of the barycentre at the $t$ epoch. The origin of the rectangular equatorial coordinates is an {\it a priori} rough estimation of the average position of the star.
}
\label{fig:wBin}
\end{figure} 

The calculation of the abscissa of the star for the observation epoch $t$ is performed in several steps, which are roughly summarised in Fig.~\ref{fig:wBin}. Hereafter, LP and LTT affects are introduced in the calculation; about the orbital delay, since it is much smaller than the period of the binary, a first order approximation is used: The eccentric anomaly $E(t)$ is computed, and the $z$-coordinate and the orbital delay $\Delta t$ are derived,
from equations (\ref{eq:z}) and (\ref{eq:Dt}). The eccentric anomaly corrected for LTT, $E(t-\Delta t)$, is then computed, and the $x$ and
$y$ coordinates in an equatorial rectangular frame are derived from equation (\ref{eq:xy}). The contribution of the orbital motion to the abscissa of the star, $w_o$, is the
projection of the ($x,y$) coordinates on the abscissa axis. When $\phi$ is the
position angle of the axis, $w_o$ is given by the expression:

\begin{equation}
w_o = x \sin \phi + y \cos \phi
\label{eq:wo}
\end{equation}

The abscissa of the star is then derived by still adding to $w_o$ the position of the barycentre, $w_B$. The
perspective effects concerning single stars are then taken into account: the parallax
is varying due to the radial motion, and the proper motion is affected by perspective
acceleration \citep{Dravins}. Therefore, the abscissa of the barycentre,
$w_B$,
is given by the equation:

\begin{equation}
\begin{array}{ll}
w_B = & \Delta \alpha_{\ast 0} \sin \phi + \Delta \delta_0 \cos \phi +
         (1-\mu_r \times ( t - \Delta t)) \times  \\ 
 & (\varpi_0 f_\varpi + \mu_{\alpha *_0} \times (t-\Delta t) \sin \phi \; + \\
 &  \mu_{\delta_0} \times (t-\Delta t) \cos \phi)
\end{array}
\label{eq:wbary}
\end{equation}

\noindent
where 
$\Delta \alpha_{\ast 0}$, $\Delta \delta_0$, $\varpi_0$, $\mu_{\alpha *_0}$, and $\mu_{\delta_0}$
are the position, parallax and proper motion of the barycentre at the epoch $t=0$, respectively. $ f_\varpi$ is the partial derivative of $w$ with respect to the actual parallax at
the observation epoch $t$. It depends on the coordinates of the binary and on the position of the Earth on its orbit. The $(1-\mu_r \times ( t - \Delta t))$ coefficient is due to barycentric perspective. The introduction of barycentric perspective in the simulated $w_B$ is necessary for being realistic, and it is
very relevant in practice: in
the next section, it will make possible the derivation of the RV with the other parameters; therefore, it contributes efficiently to the correction of the LP effect when the RV is not measured spectroscopically. It is worth noticing that the effect of the epoch correction in this term is negligible, and that it is possible to simply use $(1-\mu_r t)$.

Developing $w_o$ with equations (\ref{eq:ABFGt}) to (\ref{eq:xy}), one obtains the sum
$w_o + w_B$:

\begin{equation}
\begin{array}{ll}
w(t) = & \Delta \alpha_{\ast 0} \sin \phi + \Delta \delta_0 \cos \phi \\
 & + \;  \varpi_0 f_\varpi \; (1-\mu_r  t)\\
 & + \;  \mu_{\alpha *_0}(1-\mu_r  t -[z(t-\Delta t)]_{rad}) \; (t-\Delta t) \sin \phi  \\
 & + \;  \mu_{\delta_0} (1-\mu_r  t -[z(t-\Delta t)]_{rad}) \; (t-\Delta t) \cos \phi \\
 & + \;  A_0 \; (1 - \mu_r t)  \; X(t - \Delta t) \; \cos \phi \\
 & + \; B_0  \; (1 - \mu_r t) \; X(t - \Delta t) \; \sin \phi \\
 & + \; F_0 \; (1 - \mu_r t)  \; Y(t - \Delta t) \; \cos \phi \\
 & + \; G_0 \; (1 - \mu_r t)  \; Y(t - \Delta t) \; \sin \phi \\
\label{eq:w}
\end{array}
\end{equation}

\noindent
which is the abscissa of the photocentre when LP and LTT are both taken into account.
It is worth noticing that equations (\ref{eq:wo}) to (\ref{eq:w}) are not specific to an
observation mode, and apply to any coordinate axis. When 2-D measurements are considered, the 
ordinates obey the same formulae as the abscissae; only one parameter is changing: the position angle of the coordinate axis, $\phi$.


\subsection{Calculation of the solutions with local perspective and LTT}
\label{sec:solLP-LTT}
Thirteen parameters are necessary to entirely fit the motion of the primary component of a
binary system, {\it plus} the indication of the ascending node. In the usual order, these parameters are:
\begin{itemize}
\item
The 6 parameters of the barycentre of the binary at the reference epoch
$t=0$. These parameters are the five ones listed in Section~\ref{sec:abscissa} above (i.e. $\Delta \alpha_{\ast 0} ,
\Delta \delta_0, \varpi_0, \mu_{\alpha *_0}, \mu_{\delta_0}$), and the RV, $v_r$.
Contrarily to the parallax and the proper motion, the RV is assumed to
be constant. This parameter is optional in the calculation of the solution, since
it may be obtained independently from the astrometric data; for instance, it will be
derived from spectroscopic observations in the {\it Gaia} project. Therefore, two different
cases are considered hereafter: the case with known RV, and the case without velocity measurement, for which
$v_r$ is included among the unknowns of the astrometric solution.
\item
The Thiele-Innes elements at the reference epoch $t=0$, which are
($A_0,B_0,F_0,G_0$). Two sets of the elements $C_0$ and $H_0$ may correspond
to these 4 elements, as indicated in equations (\ref{eq:C}) and (\ref{eq:H}). For each of them, the
elements ($A,B,F,G$) are given by
equation (\ref{eq:ABFGt}) for any epoch $(t-\Delta t)$.
\item
The period, $P$, the eccentricity, $e$, and the epoch of periastron, $T_O$. These
elements are necessary to derive the eccentric anomalies, $E(t-\Delta t)$, and therefore the 3-D
coordinates of the binary components, ($x,y,z$).
\end{itemize}

These parameters are derived by $\chi^2$ minimisation, using the Levenberg-Marquardt algorithm and the routines provided by \citet{Press}. 

Each time a solution is calculated, its gaussianized GOF is derived, using the $F_2$
estimator \citep{Kendall,hip}:

\begin{equation}
F_2 = \sqrt{\frac{9n}{2}} \left[ \left( \frac{\chi^2}{n} \right)^{1/3} + \frac{2}{9n} - 1 \right]
\label{eq:F2}
\end{equation}

\noindent
where $n$ is the number of degrees of freedom.

For each virtual binary and its set of `measured' abscissae, the astrometric and orbital elements are derived three times, always taking into account the barycentric perspective effects: a first solution is derived from a simple orbital model, ignoring the local effects related to the z-dimension. LP and LTT are taken into account in the two other solutions, which are obtained assuming the right ascending node, and the
other one, respectively. The quantity $\Delta F_2$, defined as the 
difference: `GOF of the false ascending node solution'
{\it minus} `GOF of the right ascending node solution' is then derived.


\subsection{The rejection risk when local perspective and LTT are neglected}
\label{sec:rejrisk}
\begin{figure}
\includegraphics[clip=,height=2.2 in]{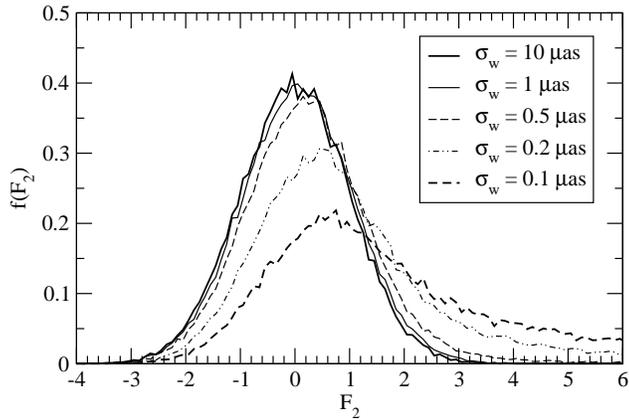}
 \caption{
The GOF of the solutions derived for the simulated binaries, when LP and LTT are neglected. The actual radial velocities are assumed for the barycentres of the systems. The distribution clearly deviates from the normal ${\cal N}(O,1)$ distribution when the error of the
astrometric measurements is less than 1~$\umu$as.
}
\label{fig:fF2PAssCor}
\end{figure} 
\begin{figure}
\includegraphics[clip=,height=2.2 in]{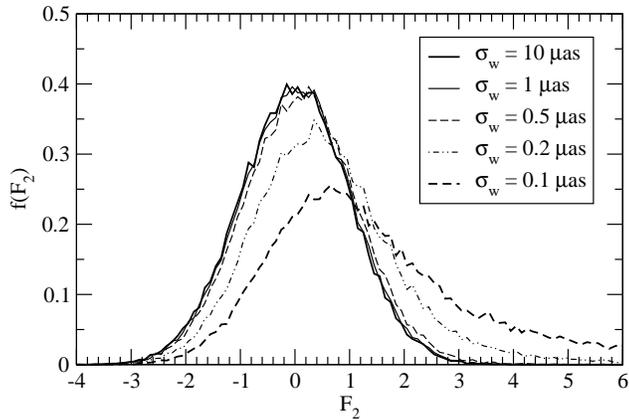}
 \caption{
Same as Fig.~\ref{fig:fF2PAssCor}, but including the radial velocity among the
unknowns of the solutions. The distribution is no more a normal ${\cal N}(O,1)$ distribution when the error of the
astrometric measurements is less than 0.5~$\umu$as.
}
\label{fig:fF2murssCor}
\end{figure}

Our aim is to estimate a limit of astrometric errors beyond which neglecting
LP and LTT would increase the rate of rejected solutions. For that
purpose, the solutions derived without LP and LTT are considered. The distributions of the GOF when the actual RV is assumed is plotted in Fig.~\ref{fig:fF2PAssCor}. Fig.~\ref{fig:fF2murssCor} refers to the solutions obtained including the RV among the unknowns.

On both figures, it appears that a 10~$\umu$as error doesn't alter the distribution
of the $F_2$ estimator, which just looks like the normal distribution ${\cal N}(O,1)$. A significant excess of systems with
$F_2 > 3$ appears only when $\sigma_w$ is less than 1~$\umu$as when the RV is known, and when $\sigma_w < 0.5$~$\umu$as otherwise.

The errors leading to visible LP or LTT effects are rather small when compared to
the estimations in Section~\ref{sec:sizeLP} and \ref{sec:sizeLTT}. The reason is
that the offsets of positions due to these effects are partly compensated by altering
the orbital elements. In fact, the modification of the orbital elements is so
efficient that the $\chi^2$ of the solution doesn't change significantly when it
is accounted for the LTT effect.
This appears when simulations are performed assuming null velocities for the barycentres of the systems, in order to suppress the LP effect. The GOF of the solutions are then always obeying the same normal distribution, no matter such effect is taken into account, and no matter the precise choice of the ascending node. 
On the other hand, the errors of the derived eccentricities and of the Campbell's angles are much larger when the wrong node is assumed in place of the right one: for instance, for the eccentricities, when $\sigma_w = 1$~$\umu$as, the ratio is already more than 2.
Therefore, taking LTT into account improves the quality of the solution, but it doesn't help in the selection of the ascending node. It is worth noticing that this conclusion applies
to unresolved binaries, and that the LTT effect is more visible in resolved binaries, as explained in
Section~\ref{sec:resolved}.


\subsection{The conditions for selecting the ascending node}

\begin{figure}
\includegraphics[clip=,height=2.2 in]{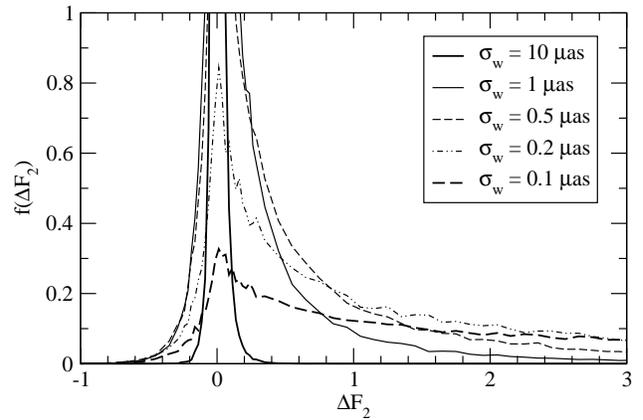}
 \caption{
Distribution of the differences between the $F_2$ estimators of the GOF when the actual radial velocity is assumed. $\Delta F_2 < 0$ means that the solution with the ascending node in the
right place leads to a $\chi^2$, and therefore a GOF, larger than the solution with the node at the opposite place.
}
\label{fig:fDGofPA}
\end{figure}
\begin{figure}
\includegraphics[clip=,height=2.2 in]{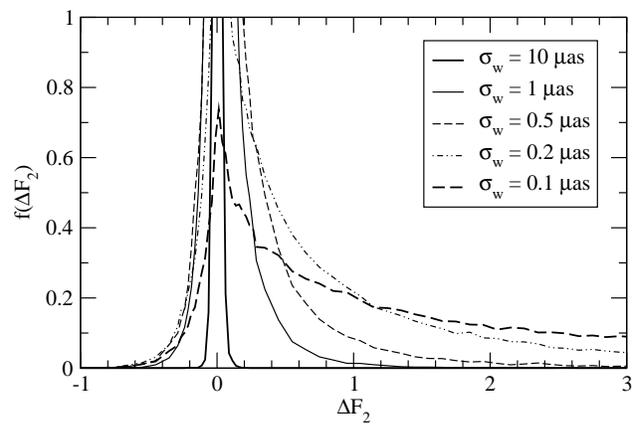}
 \caption{
Same as Fig~\ref{fig:fDGofPA}, but when then the RV is derived from the astrometric data.
}
\label{fig:fDGofmur}
\end{figure}

The differences $\Delta F_2$ defined in Section~\ref{sec:solLP-LTT} were derived from the solutions taking LP and LTT into account, and their distributions are plotted in Fig.~\ref{fig:fDGofPA} and \ref{fig:fDGofmur}. It comes from these plots that, for some systems,  $F_2$ may be used to select the ascending node even when the astrometric error is too large for providing an excess of large GOF in Fig.~\ref{fig:fF2PAssCor} and \ref{fig:fF2murssCor}. For instance, when $\sigma_w=1$~$\umu$as, the distribution of $\Delta F_2$ is clearly asymmetric, and $|\Delta F_2|>0.5$ means that
the true ascending node is very probably the one with the smallest $F_2$.


\subsection{The radial velocity derived from perspective}

\begin{figure}
\includegraphics[clip=,height=2.2 in]{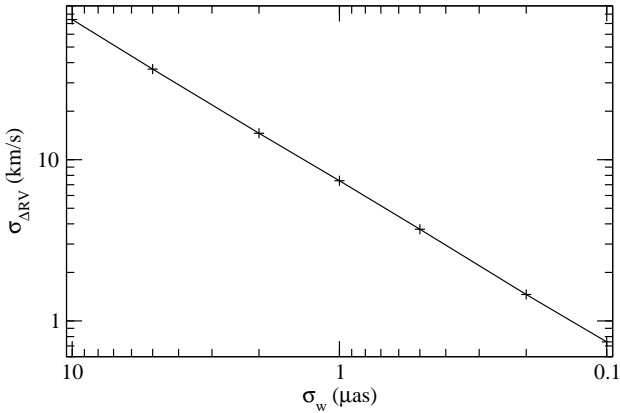}
 \caption{
Standard deviation of the radial velocity error, as a function of the error of the
astrometric measurement.
}
\label{fig:sigDRV}
\end{figure}

Not surprisingly, the solutions derived assuming the actual RV of the
barycentre are more efficient in the node selection than that including the calculation of $v_r$. Since the actual RV can never be measured in practice, the results above can only
refer to RV `accurate enough'; but what accuracy may still be considered as `enough'?
In order to answer this question, we consider the RV errors when $v_r$ is derived from
the astrometric measurements. The standard deviation of these errors, $\sigma_{\Delta RV}$, is plotted in Fig.~\ref{fig:sigDRV}, as a function of the astrometric error, $\sigma_w$. An average RV measured with an uncertainty much smaller than $\sigma_{\Delta RV}$ may be considered as equivalent to the actual one.

In the {\it Gaia} project, it is expected to get the average RV of the brightest systems, with
an accuracy of a few km/s. This is much less than the RV error when $v_r$ is derived from
the astrometric measurements, since the astrometric error of each {\it Gaia} measurement will be much larger than the upper limit of the range of $\sigma_w$ considered in Fig.~\ref{fig:sigDRV}. Therefore, the possibility to find the true ascending node with {\it Gaia} could
be obtained through the solutions assuming the actual RV. Unfortunately, since the error of the astrometric {\it Gaia} measurements at epoch (not to be confused with the average astrometric error) will be much larger than 10~$\umu$as, and it is
obvious that LP and LTT may both be ignored in the preparation of the {\it Gaia} catalogue.


\section{Application to resolved astrometric binaries}
\label{sec:resolved}


\subsection{Calculation of the astrometric solution}
\label{sec:solresolved}

The resolved astrometric binaries are less frequent than the unresolved ones, but they
are also more relevant, since the masses of the components are then directly obtained from the astrometric solution. The derivation of the parameters is not very different from the
calculation seen above: the abscissae of the primary component are again given by equation
(\ref{eq:w}), and the method described in Section~\ref{sec:solLP-LTT} may be applied with the
parameters of the primary component in place of that of the photocentre.

The abscissae of the secondary component are slightly different. It is necessary to add a
parameter to that of the primary, in order to derive the Thiele-Innes elements of the
secondary. The simplest method consists in adding the secondary semi-major axis, $a_2$. The
six Thiele-Innes elements of the secondary may then be derived from that of the primary, since $A_2 = - a_2 A_1 / a_1$, $B_2 = - a_2 B_1 / a_1$, and so on, until
$H_2 = - a_2 H_1 / a_1$ (a method for deriving $a_1$ from $A_1$, $B_1$, $F_1$ and $G_1$ is in
\citealt{Binnen}). Therefore, $X$, $Y$, $z$ and $\Delta t$ may be derived for the
secondary using the same formulae as for the primary.

When the parameters of the binary have been computed, the masses of the components are derived
from

\begin{equation}
\left\{ \begin{array}{l}
{\cal M}_1 =  a_2 \frac{(a_1 + a_2)^2}{\varpi^3 P_0^2} \\
{\cal M}_2 =  a_1 \frac{(a_1 + a_2)^2}{\varpi^3 P_0^2} 
\end{array}
\right.
\label{eq:M1M2}
\end{equation}
 
\noindent
where the masses are in solar units when $a_1$ and $a_2$ are in the same unit as $\varpi$. $P_0$ is
the period at rest, expressed in years; it is derived from the apparent period, after correction for the
Doppler effect: $P_0= P/(1+v_r/c)$. 


\subsection{Application to 2MASS J07464256 +2000321}

\subsubsection{The LP+LTT solution compared to the `no local effect' solution}

Instead of repeating the calculations performed for unresolved binaries, it is more appealing
to consider a real nearby binary, and to search when LP and LTT must be taken into account in
order to get masses as accurate as possible. The sample of nearby systems for which the orbital elements are already known is rather scarce. 2MASS J07464256 +2000321 roughly fulfils
the requirements: it is a reasonably nearby system (12 pc distant), with a period long enough
for having a wide angular separation between the component (semi-major axis of 0.2 arcsec), but short enough (10.5 y) for permitting observations covering a complete orbit. Nevertheless,
the observation frequency used in Section~\ref{sec:simulation} is adapted hereafter in order to
keep the same number of observations although the time-span is extended up to 12 years. The 
parameters of 2MASS J07464256 +2000321 are summarised in Table~1. The radial velocity comes
from \citet{Bailer}, the masses and the orbital parameters are from \citet{Bouy}; the other
data are from Simbad. 

\begin{table}
\caption{The astrometric elements of 2MASS J07464256 +2000321, as assumed in the simulations.}
\centering
\begin{minipage}{80mm}
\begin{tabular}{lc}
\hline
\multicolumn{2}{c} {2MASS J07464256 +2000321} \\
\hline
$\varpi_0$     & 81.9 mas    \\
$\mu_{\alpha*}$ & -368 mas/y  \\
$\mu_\delta$   & -39 mas/y  \\
$v_r$          & 54.1 km/s  \\
$P$            & 10.5432 y  \\
$e$            & 0.41       \\
$i$            & 141.6 $\deg$ \\
$\Omega$        & 20.7 $\deg$\footnote{$\Omega$ is known only modulo $\pi$.} \\
$\omega$       & 350.6 $\deg$  \\
${\cal M}_1$     & 0.085 ${\cal M}_\odot$  \\
${\cal M}_2$     & 0.066 ${\cal M}_\odot$  \\
\hline
\end{tabular}
\end{minipage}
\end{table}

Synthetic observations of 2MASS J07464256 +2000321 were generated, assuming a constant astrometric error for each run. The difference between the beginning of the observations
and the periastron epoch was randomly varying between 0 and $P$.

Three solutions were computed for each set of virtual observations: 
\begin{enumerate}
\item
the solution with the
ascending node in the right place, which is called the `LP+LTT solution' hereafter;
\item
the solution with the ascending node at the opposite place; this solution was
used to derive the difference $\Delta F_2$;
\item
a solution ignoring LP and LTT effects, which is referred as the `no local effect
solution'.
\end{enumerate}

\begin{table*}
 \centering
 \begin{minipage}{140mm}
\caption{The node solution and the `no local effect' solution 
for 2MASS J07464256 +2000321, for various astrometric errors.
}
\begin{tabular}{@{}lcccccccl@{}}
\hline
$\sigma_w$ & \multicolumn{3}{c} {LP+LTT solution} & \multicolumn{5}{c} {no local effect solution} \\
($\umu$as)  & $\sigma(\Delta {\cal M}_1)$ & $\sigma(\Delta {\cal M}_2)$ & $\Delta F_2 > 0$ & 
$\langle \Delta {\cal M}_1 \rangle $ & $\sigma(\Delta {\cal M}_1)$ & 
$\langle \Delta {\cal M}_2 \rangle $ & $\sigma(\Delta {\cal M}_2)$ & $\langle F2 \rangle $ \\
 & (${\cal M}_\odot$) & (${\cal M}_\odot$) & (per cent) & (${\cal M}_\odot$) & 
(${\cal M}_\odot$) & (${\cal M}_\odot$) & (${\cal M}_\odot$) &  \\
\hline
$\;$ 1 & 7.7 $10^{-7}$ & 6.1 $10^{-7}$ & 99.5 & -11  $10^{-7}$ & 22 $10^{-7}$ & -7.2 $10^{-7}$ &
 17  $10^{-7}$ & 6.8 \\
$\;$ 2  & 1.5 $10^{-6}$ & 1.2 $10^{-6}$ & 92   & -1.1  $10^{-6}$ & 2.6 $10^{-6}$ & -0.7 $10^{-6}$ &
 2.0  $10^{-6}$ & 2.1 \\
$\;$ 5 & 3.8 $10^{-6}$ & 3.1 $10^{-6}$ & 72   & -1.1  $10^{-6}$ & 4.4 $10^{-6}$ & -0.7 $10^{-6}$ &
 3.5  $10^{-6}$ & 0.37 \\
20  & 7.7 $10^{-6}$ & 6.1 $10^{-6}$ & 61   & -1.1  $10^{-6}$ & 8.0 $10^{-6}$ & -0.7 $10^{-6}$ &
 6.4  $10^{-6}$ & 0.10 \\
50  & 3.8 $10^{-5}$ & 3.1 $10^{-5}$ & 53   & -0.1  $10^{-5}$ & 3.8 $10^{-5}$ & -0.1 $10^{-5}$ &
 3.1  $10^{-5}$ & 0.01 \\
\hline
\end{tabular}
\end{minipage}
\end{table*}

For the solutions (i) and (ii), the errors of the masses, $\Delta {\cal M}_1$ and $\Delta {\cal M}_2$ were computed, as well as their standard deviations, $\sigma$. The errors are
positive when the solution provides masses larger than the actual ones, and negative
otherwise. The $F_2$ estimator of the GOF was also computed for each of the three solutions,
and the difference of $F_2$
between solution (ii) and solution (i) was derive; as in the previous section, this difference
is positive when the $\chi^2$ of the solution with the right ascending node is smaller than
that of the other node. The most relevant statistics are gathered in Table 2. 

It comes from these results that $F_2$ is much more sensitive to the local effects than
for an unresolved binary with the same semi-major axis. The reason lies in the compensation
of the LTT effect by alteration of the orbital parameters, as explained in 
Section~\ref{sec:rejrisk}. For resolved binaries, the sign of $\Delta t$ is not the same for both components; therefore, ignoring LTT results in shifting the position of the barycentre
out from
the line joining the components; no modification of the orbital parameters can reproduce such
alteration.

Another important result visible in Table~2 is the systematic error when the local effects are
neglected. The standard deviation of the mass errors are becoming similar when the astrometric error, $\sigma_w$, is around 20~$\umu$as, but the average error is still 14~per cent of the error of the
LP+LTT solution. The LP and LTT effect are really negligible only when $\sigma_w$
is around 50~$\umu$as.


\subsubsection{The `degenerate LP solution'}

Accurate masses may be computed with a LP+LTT solution, but only when the position angle
of the ascending node is known. At present, the ascending node of 2MASS J07464256 +2000321
is still not fixed despite of several radial velocity measurements, because the spectrum
of this star is a blend of the two components \citep{Bailer}. The comparison of the GOF of
the LP+LTT solutions derived from the two nodes may lead to the selection of the ascending one,
but it comes from Table~2 that this method would be reliable only when $\sigma_w$ is less than 
2~$\umu$as. Therefore, another calculation method must be used between 2 and 20~$\umu$as. This 
method takes into account a part of the LP effect, but not the effects related to the position
of the true ascending node. For that reason, its solution is called the `degenerate LP
solution' hereafter.

\begin{table}
 \centering
\caption{The mass errors and the GOF coming from the degenerate LP solution 
for 2MASS J07464256 +2000321.
}
\begin{tabular}{@{}lccccl@{}}
\hline
$\sigma_w$  & $\langle \Delta {\cal M}_1 \rangle $ & $\sigma(\Delta {\cal M}_1)$ & 
$\langle \Delta {\cal M}_2 \rangle $ & $\sigma(\Delta {\cal M}_2)$ & $\langle F2 \rangle $ \\
($\umu$as) &(${\cal M}_\odot$) & 
(${\cal M}_\odot$) & (${\cal M}_\odot$) & (${\cal M}_\odot$) &  \\
\hline
$\;$ 1  & -2.4  $10^{-7}$ & 9.4 $10^{-7}$ & -0.7 $10^{-7}$ &
 7.4  $10^{-7}$ & 0.48 \\
$\;$ 2  &  -2.4  $10^{-7}$ & 1.6 $10^{-6}$ & -0.7 $10^{-7}$ &
 1.3  $10^{-6}$ & 0.13 \\
$\;$ 5  &  -2.4  $10^{-7}$ & 3.9 $10^{-6}$ & -0.7 $10^{-7}$ &
 3.1  $10^{-6}$ & 0.03 \\
\hline
\end{tabular}
\end{table}

The degenerate LP solution takes into account the variation of the size of the orbit (i.e. of
$A_{1,2}$, $B_{1,2}$, $F_{1,2}$ and $G_{1,2}$) when the distance is changing. The calculation
consists in ignoring $z$ and $\Delta t$ in equation~(\ref{eq:w}), but not the $(1-\mu_r t)$
coefficient of the Thiele-Innes elements. The results of the degenerate LP solution are presented
in Table~3. As expected, they are quite acceptable for $\sigma_w$ larger than 2 $\umu$as, i.e.
when the ascending node cannot be determined.


\section{Summary and conclusion}
\label{sec:conclusion}

We have study the impact of two effects which have been neglected up to now in astrometric studies of binary
systems: the local perspective, and the orbital light--travel time. These effects are related to
the orientation of the orbit in space, and taking them into account may lead to the
position of the ascending node of the orbit, i.e. to the orientation of the spin vector.
The statistical distribution of the spin orientations is a clue in the study of
binary formation. It was derived and discussed in the past by
\citet{dommanget} and, more recently by \citet{glebocki}. The global distribution of the
spin looks isotropic, but Glebocki concluded from a sample of 252 orbits that subgroups seem
to have asymmetric distribution. An extension of his sample would allow to re-consider this
question.

We have found that neglecting LTT doesn't affect the $\chi^2$ of the solution for
unresolved binaries, since it
is compensated by a bias in the orbital elements, essentially
the eccentricity and the orientation of the orbit. The reason is that LTT doesn't change the shape of the apparent orbit; only the observation epochs are affected, in relation with the orbital phase. Therefore, neglecting LTT can't lead to discarding the binary model. On the other hand, taking it into account for stars with distances $d=5$~pc would significantly improve the accuracy of the orbital elements when the astrometric accuracy $\sigma_w$ is a few $\umu$as.
For other distances, the minimum $\sigma_w$ is varying as the inverse of $d$.

The ascending node of unresolved astrometric binary orbits may be found from the LP effect. However, for that purpose, astrometric measurements with accuracy around
1~$\umu$as and beyond are required. When the barycentric radial velocities are known, and therefore taken into account in the correction of perspective acceleration, LP must be taken into account when the errors of the astrometric measurements are less than the 1~$\umu$as limit. The binary star model could be erroneously discarded otherwise, since the GOF of the solution would look excessive.
When the radial velocities are not available, they are derived from the barycentric perspective effects; however the corrections for LP are then necessary only when the errors of the astrometric measurements are less than 0.5~$\umu$as.
As for LTT, the results depending on LP apply to binaries at a distance $d=5$~pc, with velocities typical of population I late-type dwarfs, and with semi-major axes of 1~AU. For other parameters, they are unchanged when the astrometric errors are varying in the same proportions as the velocities and the semi-major axes, and in the inverse proportion of $d^2$.

The resolved binaries were examined through the example of 2MASS J07464256 +2000321, a brown
dwarf system
for which all the parameters are already known, the ascending node excepted. The LTT effect
is not compensated by alteration of the orbital elements as for unresolved binaries, since it plays in opposite directions for both components.
It appeared that
neglecting LP and LTT would lead to erroneous masses when the errors of the astrometric
measurements are around 20~$\umu$as. However, when the astrometric error is larger than
1 or 2~$\umu$as acceptable evaluations of the masses may be obtained
from a `degenerate LP solution'. This calculation doesn't require the knowledge of the ascending node since it consists in correcting the semi-major axis for distance variations.
However, when the astrometric error is around 1~$\umu$as, the degenerate LP solution is no more
acceptable, and only a calculation entirely taking into account LP and LTT may lead to
masses as accurate as permitted, through the determination of the ascending node.
Since these results apply to a specific binary at a distance of 12~pc and with very light
components, it is highly probable that a few binaries in the forthcoming {\it Gaia} mission
will require a degenerate LP solution for getting accurate masses. On the other hand,
complete LP+LTT solutions will be necessary only when measurements with errors around
1~$\umu$as will be available for nearby binaries, possibly thanks to the {\it SIM} project.


\section*{Acknowledgements}

It is a pleasure to thank all the people who contributed to this work by comments and discussion. I
am grateful to Fr\'ed\'eric Arenou, Mustapha Mouhcine and Dimitri Pourbaix for reading and correcting the manuscript, but also to Guillem Anglada-Escud\'e and Thierry Forveille for stimulating comments, to Ulrich Bastian and Luc Blanchet for helpful explanations, and to Lennart Lindegren for checking some formulae.

An anonymous referee read carefully the manuscript and made constructive comments, inducing many improvements
in the presentations of the methods and of the calculations. 

The Simbad database of the CDS was used to get relevant informations about
2MASS J07464256 +2000321.



\end{document}